# Spin-phonon coupling suppressing the structural transition in perovskite-like oxide

Shalini Badola[1], Supratik Mukherjee[2], Greeshma Sunil[1], B. Ghosh[1], Devesh Negi[1], G. Vaitheeswaran[2*], A. C. Garcia-Castro[3*], and Surajit Saha[1*]

[1]*Indian Institute of Science Education and Research Bhopal, Bhopal 462066, India.*

[2]*School of Physics, University of Hyderabad, Prof. C. R. Rao Road, Gachibowli, Hyderabad 500046, Telangana, India.*

[3]*School of Physics, Universidad Industrial de Santander, Calle 09 Carrera 27, Bucaramanga, Santander, 680002, Colombia.*

*Correspondence: surajit@iiserb.ac.in, acgarcia@uis.edu.co, vaithee@uohyd.ac.in

**Abstract:** Multifunctional properties in quantum systems require the interaction between different degrees of freedom. As such, spin-phonon coupling emerges as an ideal mechanism to tune multiferroicity, magnetism, and magnetoelectric response. In this letter, we demonstrate and explain, based on theoretical and experimental analyses, an unusual manifestation of spin-phonon coupling, *i.e.,* prevention of a ferroelastic structural transition, and locking of high-temperature *R-3m* phase in a magnetically frustrated perovskite-like oxide $Ba_2NiTeO_6$. We present $Ba_2NiTeO_6$ as a prototype example among its family where long-range antiferromagnetic structure couples with a low-frequency $E_g$ mode (at 55 cm$^{-1}$) that exhibits a large anharmonicity. Our findings establish that spin-phonon coupling clearly suppresses the phonon anharmonicity preventing the structural phase transition from the *R-3m* to the *C2/m* phase in $Ba_2NiTeO_6$.

Non-linear interaction between spins and phonons is critical not only for application-oriented phenomena like multiferroicity, magnetoelectricity, superconductivity, etc. but to stabilize or melt specific magnetic and structural phases as well [1, 7]. This has invigorated intense investigations on the cross-coupling between electron, spin, and phonon degrees of freedom in the realm of recent condensed matter research [8-11]. Spin-phonon coupling (SPC) provides a spin-relaxation channel to exchange energy between spin excitation and the thermal bath of the lattice [12]. Broadly, two fundamental mechanisms describe SPC in transition metal oxides. The conventional approach explains SPC mediating via exchange pathway ($J$) modulated by dynamical bond distance or bond angle during the vibration of the phonon mode [13]. Alternatively, it is understood through single-ion anisotropy that dominates in highly spin-orbit coupled $4d/5d$ systems [14]. Magnetic frustration is well-known to further catalyze this coupling [15]. Thus, SPC manifests into physical phenomena that involve predominantly magnetism. Historically, spin-phonon coupling has been known to promote structural transition/distortion, especially when there is magnetic frustration [4, 16-19]. Moreover, in $SrMnO_3$, for example, SPC is responsible for the emergence of strain-induced multiferroicity [20-22]. Nevertheless, there are no studies so far, to the best of our knowledge, that establish SPC as a potential mechanism to prevent structural transition in transition metal oxides.

Perovskite $Ba_2NiTeO_6$ with buckled honeycomb-type spin-lattice offers an excellent platform to investigate SPC and its effect on the structural properties. The unique bilayer triangular arrangement of spins results in competing $J$'s that induce strong magnetic frustration which eventually gets relieved to an antiferromagnetic (AFM) state below $T_N \sim 8.6$ K [23, 24]. Unlike conventional transition metal oxides, $Ba_2NiTeO_6$ exhibits an unusual long-chain super-exchange mediated via multiple ions in between the two nearest magnetic ions ($Ni^{2+}$). Consequently, SPC is expected to be extremely weak to cause any significant effect on the lattice dynamics. $Ba_2NiTeO_6$ crystallizes into trigonal symmetry ($R$-$3m$) that remains stable down to ~2 K [24, 26]. On the contrary, its isostructural non-magnetic counterpart $Ba_2ZnTeO_6$ undergoes a concomitant ferroelastic and structural transition to monoclinic phase ($C2/m$) below ~ 150 K due to soft mode instability arising from strong anharmonicity [27, 28]. Thus, a distinction in the two structural dynamics despite their close similarities has remained a puzzle to date and needs to be addressed.

In this letter, we demonstrate the cross-talk between magnetic and structural dynamics of $Ba_2NiTeO_6$ using Raman scattering, X-ray, and magnetic measurements as well as first-principles calculations. We find that most of the phonon modes deviate from the expected cubic-anharmonic trend below ~ 100 K where spin-spin short-ranged interaction emerges, indicating the onset of SPC. The SPC reduces the anharmonicity of the phonon mode at ~ 55 cm$^{-1}$ (which otherwise is a soft mode in Ba$_2$ZnTeO$_6$) and magnetic frustration in Ba$_2$NiTeO$_6$, thereby preventing the *R-3m* to *C2/m* structural transition [28].

All the experiments were performed on polycrystalline pellets of Ba$_2$NiTeO$_6$ and Ba$_2$ZnTeO$_6$ grown using solid-state synthesis method as described in Supplemental Material [29]. The experimentally recorded and Density Functional Theory (DFT)-calculated X-ray diffraction pattern of Ba$_2$NiTeO$_6$, shown in Fig. 1(a), confirm the trigonal crystal structure. The evolution of the lattice parameters with temperature was recorded by a PANalytical X-ray diffractometer attached to an Anton Paar TTK 450 heating stage, as shown in Fig. 1(b). Temperature evolution of the phonon modes was captured using an HR Evolution Raman spectrometer ($\lambda_{laser}$ = 532 nm) coupled to an attoDRY 1000/SU cryostat and a linkam heating stage (Model HFS600E-PB4). Magnetic measurements were performed using Quantum Design SQUID-VSM magnetometer. The experimental results were verified using first-principles calculations implemented in the VASP code [30-39]. The methods used are further detailed in the Supplemental Material [29].

In general, various factors contribute to phonon anomalies over temperature (*T*) including spin-phonon, phonon-phonon, and electron-phonon couplings [40, 41]. We analyze the *T*-dependence of phonon frequencies (ω) illustrated in Fig. 2(a) using standard cubic-anharmonic formalism given by [41]

$$\Delta\omega^{canh}(T) = \omega_0 + A\left[1 + \frac{2}{e^{x(T)}-1}\right] \quad (1)$$

where $x(T) = \frac{\hbar\omega}{2k_BT}$ with ℏ as reduced Planck's constant and $k_B$ as the Boltzmann coefficient. Moreover, $\omega_0$ and $A$ are the fitting parameters. $\Delta\omega^{canh}(T)$ expects phonon frequency to decrease with progressing temperature. On the contrary, our analysis reveals that phonon frequencies in Ba$_2$NiTeO$_6$ deviate from $\Delta\omega^{canh}(T)$ at temperatures below ~ 100 K. The insulating nature of the system excludes the possible contribution from electron-phonon coupling to cause this deviation. Therefore, we explore other possibilities to address the observed phonon anomalies.

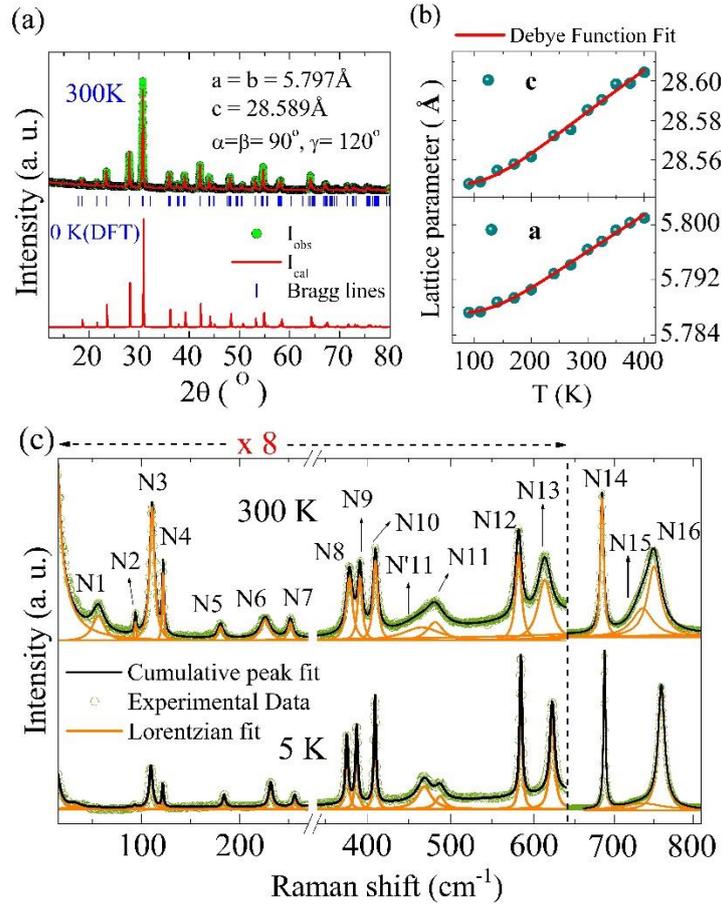

**Figure 1** (Color online): (a) DFT calculated and room-temperature X-ray diffraction profiles of Ba$_2$NiTeO$_6$ displaying the lattice parameters (b) Temperature evolution of the lattice parameters fitted (solid red line) using Debye equation (Supplemental Material [29]) (c) Raman spectra of Ba$_2$NiTeO$_6$ at a few temperatures showing deconvolutions of phonons as N1 to N16 and N'11 at 300 K.

Previous reports and our data suggest that Ba$_2$NiTeO$_6$ demonstrates a variety of magnetic properties, including magnetic frustration T$_N$ < T < |θ$_{CW}$|), a short-ranged ordering (at ~ 35 K), and an AFM order below T$_N$ (~ 8.6 K) [23, 24]. A careful investigation of the magnetization data, displayed in Fig. 2(c), suggests that similar to the phonons, χ$^{-1}$ also deviates from Curie-Weiss behavior below ~100 K, implying a reduction in magnetic frustration and the onset of spin-spin correlation. We note that the temperature regime of phonon and magnetic anomalies largely coincides with each other. Therefore, temperature-dependent phonon anomalies below ~100 K can be attributed to spin-phonon coupling. Since spin-spin correlations exist for this system even above the T$_N$, we combine Cottom and Lockwood's models along with the conventional Mean-Field approach to quantify the SPC constants (λ) (see Supplemental

**Table I**. The spin-phonon coupling coefficients (λ) for the phonon modes in $Ba_2NiTeO_6$.

| Phonons | λ (cm$^{-1}$) | | Phonons | λ (cm$^{-1}$) | |
|---|---|---|---|---|---|
| | Experiment | DFT | | Experiment | DFT |
| N1 | * | 34 | N9 | * | 0.1 |
| N2 | 0.7 | 0.1 | N10 | 0.9 | * |
| N3 | 0.8 | 1.6 | N11 | 1.5 | 0.7 |
| N4 | 0.6 | 0.1 | N12 | 1.3 | 0.1 |
| N5 | 0.6 | 0.2 | N13 | 0.2 | 0.1 |
| N6 | 0.5 | 1.3 | N14 | 1.1 | 5.3 |
| N7 | 0.8 | * | N15 | 5.1 | 5.7 |
| N8 | * | 0.9 | N16 | 0.5 | * |

*: Refer to text and supplemental material for more details

Material Ref. [29]) [41 - 45], which are found to be comparable to those in $MnF_2$, $HoCrO_3$, $Ca_2RuO_4$, $Fe_3GeTe_2$, and $Ni_2NbBO_6$ mainly lying in the range of 0.3 – 1.5 cm$^{-1}$, as listed in Table I [43, 46-49]. It may be noted that the modes N1, N8, and N9 exhibit anomalous softening upon cooling for the entire temperature range in our experiment. Therefore, these models may not be suitable for an accurate estimation of the values of λ for these modes. More details on the analysis of the spin-phonon coupling for N1, N8, and N9 modes can be found in Section V of Supplemental Material [29]. We believe that such a coupling arises from an indirect super-exchange interaction ($Ni^{2+}$-$O^{2-}$-$Te^{6+}$-$O^{2-}$-$Ni^{2+}$) mediated via corner- and/or face-shared $TeO_6$ octahedra, as discussed later.

Unlike other phonon modes, the cubic-anharmonic model could not explain the behavior of N1, N8, and N9 modes since their frequency decreases upon cooling in the entire temperature range (see Fig. 2(a) and 2(b)). In particular, the lowest frequency phonon mode N1 exhibits a large softening of $\frac{\Delta\omega}{\omega} = \frac{\omega(5K) - \omega(300K)}{\omega(5K)} \sim$ - 70 % upon cooling, as opposed to a typical hardening of ~ 1-2 % (see section IV in Supplemental Material for more details [29]). The fact that the contribution from SPC manifests only below intermediate temperatures (~ 100 K) and the electron-phonon coupling is absent, we ascribe the anomalous large softening of phonon mode (N1) upon cooling (T > 100 K) to strong anharmonicity. In general, the total lattice anharmonicity ($\Delta\omega^{canh}(T)$) carries the contribution from both quasi-harmonic and intrinsic anharmonic components. We estimate the quasi-harmonic contribution to be small (~ 0.7 %) for N1 from the temperature-dependent X-ray diffraction (see Supplemental Material

[29]). This implies that a strong intrinsic phonon anharmonicity exists in $Ba_2NiTeO_6$ but does not lead to a structural transition like the one in $Ba_2ZnTeO_6$.

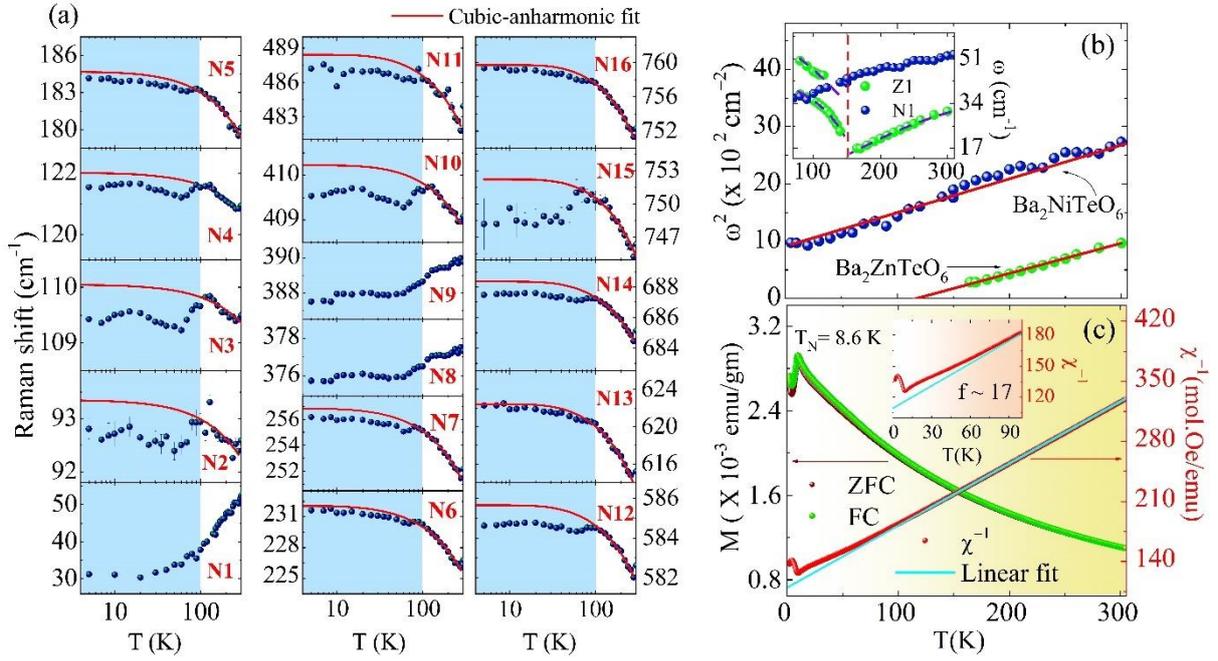

**Figure 2** (Color online): (a) Phonon frequency as a function of temperature showing deviation from cubic-anharmonic trend (solid red line) below 100 K. Error bars are included to show standard deviation in phonon frequencies. (b) Square of frequency ($\omega$) of phonon mode (N1/Z1) with temperature in isostructural systems $Ba_2NiTeO_6$ and $Ba_2ZnTeO_6$. Solid (red) lines are fit to Cochran's relation $\omega^2 = A (T - T_c)$. Inset shows the thermal response of phonon frequency of mode N1/Z1 across the transition temperature (shown with a vertical dashed red line) of $Ba_2ZnTeO_6$. (c) Temperature-dependent magnetization (M) and susceptibility ($\chi$) of $Ba_2NiTeO_6$ showing $T_N \sim 8.6$ K. Inset shows $\chi^{-1}$ vs. T representing an onset of deviation from the Curie-Weiss linear fit below 100 K

Unprecedented softening is typical of soft modes in the vicinity of structural transition due to lattice anharmonicity. Similar to the soft mode Z1 ($E_g$) at ~ 30 cm$^{-1}$ in $Ba_2ZnTeO_6$, we have modelled the phonon N1 (~ 55 cm$^{-1}$) in $Ba_2NiTeO_6$ with Cochran's relation: $\omega^2 = A (T - T_c)$, as shown in Fig. 2(b) [27, 28]. We find that Cochran's fit to N1 phonon frequency does not tend to zero at any finite temperature, unlike the mode Z1 in $Ba_2ZnTeO_6$. Therefore, N1 does not qualify to be a soft mode despite exhibiting a large softening. All these observations suggest that the phonon-phonon interactions may not be sufficient to drive a structural transition in $Ba_2NiTeO_6$, and it is important to consider the effects of other competing interactions (such as the active magnetic response).

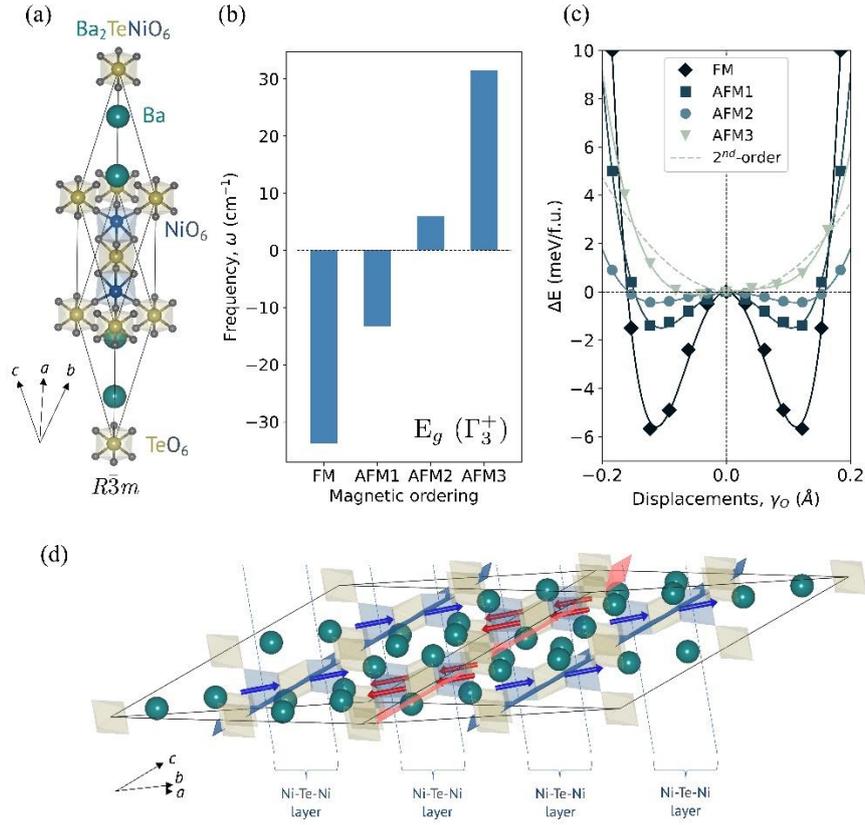

**Figure 3** (Color online): (a) Primitive (1X1X1) structural unit cell of Ba$_2$NiTeO$_6$ without considering the magnetic ordering (b) Calculated frequencies of the $\Gamma_3^+$ mode (N1) for different magnetic (FM, AFM1, AFM2, and AFM3) orderings with different propagation *q*-vectors. The unstable, (*i.e.*, imaginary) frequencies are shown as negative values by notation. (c) Total energy-wells obtained as a function of the rotational oxygen frozen displacements, $\gamma_o$ (*i.e.*, condensation of the eigen displacements associated with the N1 mode) and computed at different magnetic orderings. The latter total energy values are shown with reference to the undistorted structure. The displacements are measured as the averaged distortion of the oxygen sites involved in the octahedral rotation. (d) (2X2X2) supercell in which the corner-shared TeO$_6$ octahedra couples magnetism and phonon anharmonicity. The red and blue planes represent the magnetic planes in the AFM3 ordering. Additionally, the ferromagnetic NiO$_6$-TeO$_6$-NiO$_6$ path can be observed in the magnetic plane meanwhile the antiferromagnetic interactions are mediated by the TeO$_6$ octahedra.

In order to obtain an understanding of the distinction between the lattice dynamics of Ba$_2$NiTeO$_6$ and Ba$_2$ZnTeO$_6$, we made use of first-principles calculations, in the framework of density-functional theory, including different magnetic configurations. After full electronic and structural relaxation, the lattice parameters are found to be $a = b = c = 10.072$ Å and $\alpha = \beta = \gamma$

= 33.39° in the trigonal primitive cell reference (see Fig. 3(a)). These are equivalent to $a = b = 5.787$ Å, $c = 28.504$ Å, and $α = β = 90.0°$ and $γ = 120.0°$ in the hexagonal unit cell reference and in full agreement with our experimentally measured values. In order to compute the coupling between the lattice dynamics and the magnetic structure, we have explored possible magnetic ground states in the $Ba_2NiTeO_6$. The ferromagnetic (FM) ordering was considered initially for reference followed by three different antiferromagnetic orderings, labeled as AFM1, AFM2, and AFM3, see Supplemental Material [29]. In the AFM1 state, all the first neighbours are aligned antiferromagnetically and the propagation $q$-vector can be represented as $q = (0, 0, 0)$ in the trigonal primitive cell (*i.e.*, 20-atoms reference). In the AFM2 phase, the Ni-Ni interaction is ferromagnetic along Ni-Te-Ni face-shared octahedra (*z*-axis) for the first two layers whereas the third layer is aligned antiferromagnetically. It can again be represented by a $q = (0, 0, 0)$ in the hexagonal unit-cell (*i.e.*, 60-atoms reference). Finally, the AFM3 state (*akin* to AFM2) shows a ferromagnetic alignment along the *z*-axis in the Ni-Te-Ni octahedra. However, the overall magnetic structure is AFM along the [1,1,0] family of planes, as shown in the Supplemental Material (see Fig. S4(e)) [29]. As such, the AFM3 ordering can be considered with a $q = (1,1,0)$ magnetic propagation vector with respect to the (2x2x2) primitive unit cell reference and it is equivalent to the (0,0.5,1.0) $q$-vector in the (1x2x1) hexagonal unit cell reference, consistent with the previous report by Asai *et al.* based on neutron diffraction analysis [24]. The latter cells and orderings are shown in Fig. S4 in the Supplemental Material [29]. The full structural relaxation under all of the above magnetic orderings indicates that the AFM3 has the lowest energy configuration, agreeing to the magnetic ground state predicted in Ref. [24].

After defining various magnetic ground states, we estimated the phonon frequencies at the Γ-point considering each magnetic configuration. Interestingly, we obtained an unstable (*i.e.*, negative frequency value by notation) phonon mode (N1) at 0 K associated with $TeO_6$ octahedral rotation with $Γ_3^+$ irreducible representation. In the FM state, the phonon mode N1 shows a negative frequency of $ω_{Eg} = -44$ cm$^{-1}$. Surprisingly, when the antiferromagnetic orderings (AFM1, AFM2, and AFM3) and larger magnetic cells are considered, the unstable phonon frequency is reduced and becomes stable, $ω_{Eg} = 31$ cm$^{-1}$, for AFM3, as presented in Fig. 3(b). These observations confirm the strong correlation between phonon anharmonicity and magnetism in $Ba_2NiTeO_6$. The phonon mode (N1) is also observed in the isostructural compound $Ba_2ZnTeO_6$ in our previous work (labeled as Z1) [28] and is found to be responsible for the structural transition from *R-3m* to the *C2/m* crystal symmetry when the temperature is lowered. Nonetheless, as commented before, we do not observe the structural transition in our experiments and theoretical calculations for $Ba_2NiTeO_6$.

In order to establish coupling between phonon anharmonicity and magnetism, as indicated above, we have condensed or frozen the $E_g$ (*i.e.*, $\Gamma_3^+$) mode by including the eigen displacements, associated with the phonon mode, into the high-symmetry *R-3m* phase. As expected from the values of the unstable modes, the gain in energy in the energy-well profile is considerably reduced when larger magnetic cells are considered up to a point at which we obtain a single energy well for AFM3, indicating no gain in energy and a high-symmetry stable phase, as shown in Fig. 3(c). As in the FM, AFM1, and AFM2 cases, the energy well (*i.e.*, $\Delta E$ *vs.* $\gamma_o$) for the AFM3 phase is better fitted with the relations up to quartic terms. Further, to highlight the deviation of the obtained curve, we have also included the second-order fitting considering quasi-harmonic approximation. We find the quasi-harmonic fitting relation as $\Delta E = (105.5\gamma^2 - 2.7\gamma)$ in contrast to the quartic relation $\Delta E = (4920.2\gamma^4 - 288.8\gamma^3 - 7.7\gamma^2 + 3.0\gamma)$. Note that the pure harmonic approximation implies $\Delta E \propto \gamma^2$. This highlights the importance of the quartic 4th-order terms supporting the strong anharmonicity of the $\Gamma_3^+$ mode in $Ba_2NiTeO_6$, as also observed in $VO_2$ [50] and $ScF_3$ [51]. As such, these results clearly indicate that the interactions between the spins and the lattice prevent the *R-3m* to *C2/m* phase transition in $Ba_2NiTeO_6$, as opposed to the one observed for the isostructural compound [28]. As a computational experiment, we also relaxed and computed the phonon modes in the non-magnetic (NM) phase at 0 K. As expected, the $\Gamma_3^+$ is strongly unstable with a negative frequency value of -385 cm$^{-1}$ confirming that magnetic interactions are crucial to lock-in the *R-3m* phase through spin-phonon coupling in $Ba_2NiTeO_6$. Moreover, in terms of geometrical considerations and tolerance factor, the only (negligible)difference between the two compounds relies upon the atomic radius of Ni and Zn, which in the case of 2+ oxidation state is almost identical with $r_{Ni}$ = 87 pm and $r_{Zn}$ = 88 pm, respectively [52]. This further ascertains spin-phonon coupling to be responsible for preventing the structural transition in $Ba_2NiTeO_6$.

We have included the calculated Raman frequencies for perovskite $Ba_2NiTeO_6$ considering the AFM3 magnetic ground state in Supplemental Material [29]. The frequency values are in good agreement with our experimental Raman data. Moreover, the symmetry-mode assignment as well as the atomic species and atomic displacements involved are included for each mode. We have also estimated the spin-phonon coupling parameters ($\lambda$) for each of the modes, as listed in Table I (More details in Supplemental Material [29]). To computationally estimate the paramagnetic reference, we have taken the averaged FM, NM, AFM1, AFM2, and AFM3 frequency values as performed before for the $MnF_2$, $ABF_3$, and $Mn_3NiN$ compounds [53-55]. Since different approaches are used for the estimation of λ in experiment and theory, it is likely to give a difference in their values which is considerable.

Finally, our experiments and calculations indicate that the rotation of the corner-shared $TeO_6$ octahedra (along with Ba/Ni(Zn) displacements), involved in the lowest frequency phonon (N1(Z1)), produces unprecedented anharmonicity. Interestingly, the same $TeO_6$ octahedra connect the antiferromagnetically aligned adjacent face-shared $NiO_6$-$TeO_6$-$NiO_6$ layers (Ni-Te-Ni Layers) of $Ba_2NiTeO_6$ and mediate the super-exchange interaction *J* through $Ni^{2+}$-$O^{2-}$-$Te^{6+}$-$O^{2-}$-$Ni^{2+}$ pathway to favor spin-phonon coupling (see Fig. 3(d)). In this way, the structural stability of *R-3m* phase is ensured through entanglement(coupling) with the magnetic response in $Ba_2NiTeO_6$. The mediation of the spin-phonon coupling for the anomalous N8 and N9 modes is discussed in Supplemental Material [29].

To conclude, an unconventional mechanism has been identified to prevent structural transition in $Ba_2NiTeO_6$ using temperature-dependent Raman spectroscopy, X-ray, and magnetic measurements as well as first-principles calculations. We observe phonon anomalies to appear well above the $T_N$ (~ 8.6 K) wherein short-ranged magnetic interactions initiate (below ~ 100 K), depicting the onset of spin-phonon coupling. Our calculations establish the rare phenomena that spin-phonon coupling suppresses the strong phonon anharmonicity and magnetic frustration in $Ba_2NiTeO_6$, thereby, preventing the *R-3m* to *C2/m* structural transition (unlike in $Ba_2ZnTeO_6$). We emphasize that exploring such coupled structural and magnetic phenomena (spin-phonon coupling) that can control phase transition is important to induce multiferroic or magnetostrictive properties in non-multiferroic systems thus opening up new avenues.

## ACKNOWLEDGEMENTS

SS acknowledges Science and Engineering Research Board (SERB) for funding through ECR/2016/001376 and CRG/2019/002668. Funding from DST-FIST (Project No. SR/FST/PSI-195/2014(C)), Nano-mission (Project No. SR/NM/NS-84/2016(C)) and Ministry of Education (Grant No. STARS/APR2019/PS/662/FS) are also acknowledged. D.N. acknowledges the fellowship (09/1020(0139)/2018-EMR-I) support from CSIR. Authors acknowledge Central Instrumentation Facility at IISER Bhopal for providing access to temperature-dependent XRD and SQUID-VSM facilities. Calculations presented in this paper were carried out using the GridUIS-2 experimental testbed, being developed under the Universidad Industrial de Santander (SC3-UIS) High Performance and Scientific Computing Centre, development action with support from UIS Vicerrectoría de Investigación-y Extensión (VIE-UIS) and several UIS research groups as well as other funding resources. S.M acknowledge the financial support from DRDO, India, via ACRHEM (DRDO/18/1801/2016/01038: ACRHEM-PHASE-III). G.V. would like to acknowledge Institute of Eminence, University of Hyderabad (UOH-IOE-RC3-21-046) for the financial support and CMSD, University of Hyderabad for providing Computational facility.

# Supplemental Material

# Spin-phonon coupling suppressing the structural transition in perovskite-like oxide

Shalini Badola[1], Supratik Mukherjee[2], Greeshma Sunil[1], B. Ghosh[1], Devesh Negi[1], G. Vaitheeswaran[2*], A. C. Garcia-Castro[3*], and Surajit Saha[1*]

[1]*Indian Institute of Science Education and Research Bhopal, Bhopal 462066, India.*

[2]*School of Physics, University of Hyderabad, Prof. C. R. Rao Road, Gachibowli, Hyderabad 500046, Telangana, India.*

[3]*School of Physics, Universidad Industrial de Santander, Calle 09 Carrera 27, Bucaramanga, Santander, 680002, Colombia.*

This supplemental material contains details of the analysis of interatomic bond distances using X-ray diffraction, estimation of quasi-harmonic term and spin-phonon coupling constants, magnetic-susceptibility ($\chi^{-1}$- T), calculated magnetic configurations, and phonon symmetry assignments in $Ba_2NiTeO_6$.

# I. EXPERIMENTAL AND THEORETICAL METHODS

## A. Experimental Details:

Polycrystals of $Ba_2NiTeO_6$ and $Ba_2ZnTeO_6$ were prepared using a solid-state reaction synthesis route with high purity oxides $BaCO_3$, NiO, ZnO, and $TeO_2$ (Sigma-Aldrich). The oxides were mixed and treated at 750 °C for 4 hrs to ensure complete conversion of tetravalent tellurium ions into their hexavalent oxidation state and then subsequently calcined at 900 °C for 6 hrs with intermittent grinding after each thermal treatment. Finally, the pelletized sample was sintered at 1050 °C in air for 6 hrs. Crystalline quality and phase purity of the polycrystals were confirmed by X-Ray Diffraction (XRD) technique using the PANalytical X-Ray diffractometer. An Anton Paar 450 TTK stage was used to perform temperature-dependent XRD measurements. Energy Dispersive X-Ray Analysis (EDAX) measurements were performed at multiple spots on the sample to examine the elemental composition thus confirming the stoichiometry. Further, Raman spectroscopic measurements were performed in back-scattering geometry using the Jobin-Yvon LABRAM-HR Evolution Raman spectrometer coupled with a Peltier-cooled CCD detector. The sample was illuminated using Nd:YAG laser beam of wavelength 532 nm guided through a 50X objective with an incident power of ~ 1mW. Temperature-dependent Raman measurements were performed in two steps: Firstly, in the paramagnetic region from 80 to 600 K using the Linkam heating stage. Secondly, to explore the phonon properties in magnetically ordered phase down to 5 K, a closed-cycle attoDRY 1000/SU cryostat was interfaced with the Raman spectrometer for the measurements. Since we do not see any anomalies in the thermal response above 300 K, the Raman data are shown between 5-300 K measured using attoDRY. Raman spectra with the attoDRY system were measured in two different configurations using: (1) optical fiber, and (2) free beam optics. The bandpass filter used in optical fiber configuration restricted the spectra down to 50 cm−1. Hence, the N1 mode was separately measured using free-beam configuration but at different temperature steps. The magnetization of the sample was measured using Quantum Design SQUID-VSM with an applied magnetic field of 200 Oe.

## B. Theoretical and Computational Approaches:

We have performed density-functional theoretical [1, 2] calculations as implemented in the VASP code (version 5.4.4) [3, 4]. In the projected-augmented waves, PAW [5], the approach was used to represent the valence and core electrons. The electronic configurations considered in the pseudo-potentials for the calculations were Ba:($5s^2 5p^6 6s^2$, version 06Sep2000),

Ni:($4s^23d^8$ version 06Sep2000), Te: ($5s^25p^4$ version 08Apr2002), and O: ($2s^22p^4$, version 08Apr2002). The exchange-correlation was represented within the generalized gradient approximation GGA-PBEsol parametrization [6] and, due to strong exchange-correlation, we used the PBEsol+U [7] approach with a value of U = 4.0 eV in the 3d:Ni states. The periodic solution of the crystal was represented by using Bloch states with a Monkhorst-Pack [8] k-point mesh of 7×7×7, in the trigonal primitive cell, and 600 eV energy cut-off to give forces convergence within the error smaller than 0.001 eV.Å$^{-1}$. Born effective charges were obtained within the density functional perturbation theory (DFPT) [9] as implemented in the vasp code whereas the phonon-dispersions were computed by means of the finite-displacements method [10, 11]. Phonon dispersions were post-processed in the Phonopy code [12]. The atomic structure figures were elaborated with the support of the vesta [13].

## II. VARIATION OF THE LATTICE PARAMETERS AND BOND DISTANCES OVER TEMPERATURE

The lattice parameters (a and c), shown in Fig.1 (main text), as a function of temperature are fitted using the Debye function [14]:

$$a(T) = a_0 \left[ 1 + \frac{b\, e^{\frac{d}{T}}}{T(e^{\frac{d}{T}}-1)^2} \right] , \quad c(T) = c_0 \left[ 1 + \frac{m\, e^{\frac{n}{T}}}{T(e^{\frac{n}{T}}-1)^2} \right] \tag{1}$$

where $a_0$, $c_0$, b, d, m, and n are the fitting parameters. A fitting of lattice parameters using Eqn. 1 yield $a_0$= 5.7872 ± 0.0003 Å, b = 3.7218 ± 0.5936 K, d = 688.85 ± 46.43 K, $c_0$= 28.547 ± 0.002 Å, m = 2.7709 ± 0.6504 K, and n = 660.54 ± 66.89 K. Since we do not observe any anomaly in the thermal behavior of the lattice parameters and volume (refer main text and Fig. S1), it can be concluded that the crystal lattice of Ba$_2$NiTeO$_6$ follows the conventional (positive) thermal expansion. Moreover, the change in volume is estimated to be 0.7 % over a temperature range of 90 - 400 K (Fig. S1(a)). Though the bond distances, shown in Fig. S1(b), present some fluctuations/irregularities at low temperatures below 200 K, they do not lead to any crystallographic transition to a new structure. It is to be noted, based on the magnetization data, that the region of magnetic frustration in Ba$_2$NiTeO$_6$ lies in between 8.6 K ($T_N$) < T < 150 K (|θcw|), that intriguingly coincides with the temperature range where fluctuations/irregularities in bond distances are also observed. This is suggestive of a possibility that the lattice instability (fluctuations in bond length) appearing at low temperatures may be associated with the magnetic frustration. Though a clear structural phase transition is absent in $Ba_2NiTeO_6$ down to very low temperatures (∼ 2 K), as can be noted from our data and according to a previous report [15, 16], a possible presence of local lattice deformations/distortions triggered by magnetic frustration cannot be completely ruled out. Further experiments, such as, neutron/muon scattering to probe local lattice environments are required to confirm this possible correlation which is beyond the scope of this work.

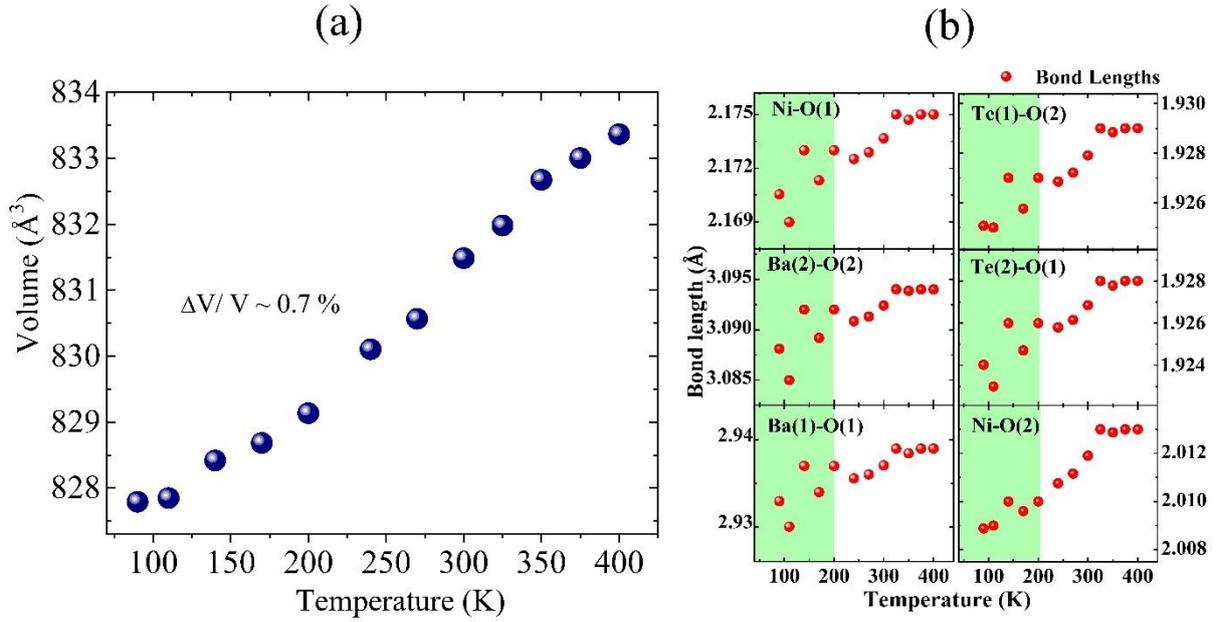

**Fig. S1**: (a) Variation of the unit-cell volume over temperature (b) Temperature-dependent bond distances showing fluctuations below 200 K.

### III. $\chi^{-1}(T)$: MAGNETIC FRUSTRATION VS LATTICE FLUCTUATIONS

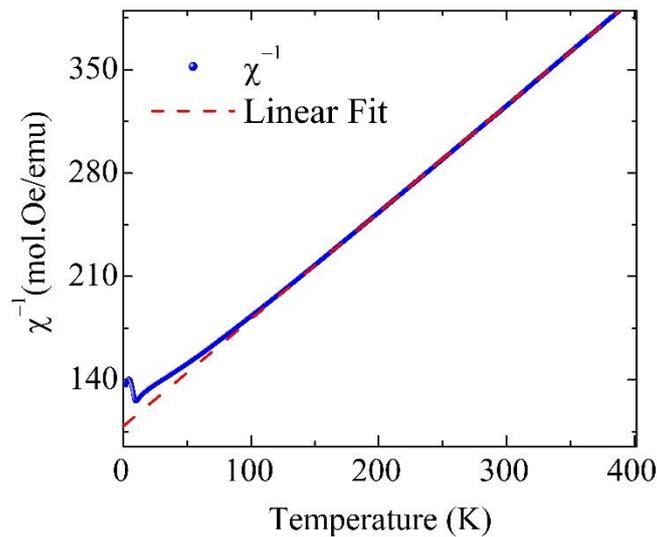

**Fig. S2**: Inverse susceptibility of $Ba_2NiTeO_6$ as a function of temperature showing the onset of deviation from the Curie-Weiss linear fit below 150 K.

In order to identify the region of magnetic frustration, inverse magnetic susceptibility is shown with the variation of temperature. A Curie-Weiss linear fit to the $\chi^{-1}$, shown in Fig. S2, shows the appearance of deviation from the linear trend below ~ 150 K indicating an onset of magnetic frustration in $Ba_2NiTeO_6$. An onset of frustration corroborates well with the fluctuations observed in the bond lengths extracted from the temperature-dependent X-ray diffraction (see Fig. S1(b)). Further, it is observed that the signatures of spin-phonon coupling become more prominent in the region where inverse magnetic susceptibility deviates from the linear trend (as discussed in the main text). These observations suggest the presence of a relation among the magnetic frustration, bond length fluctuations, and spin-phonon coupling, all arising simultaneously with the onset of short-ranged spin-spin correlations.

## IV. ESTIMATION OF QUASI-HARMONIC FREQUENCY SHIFT

As discussed in the main text, the renormalization of phonon frequency upon variation of temperature is expressed as [17]

$$\omega(T) = \omega(0) + \Delta\omega^{qh}(T) + \Delta\omega^{anh}(T) + \Delta\omega^{el-ph}(T) + \Delta\omega^{sp-ph}(T) \qquad [2]$$

where $\omega(0)$ is the phonon frequency at 0 K, $\Delta\omega^{qh}(T)$ represents the quasi-harmonic term, $\Delta\omega^{anh}(T)$ and $\Delta\omega^{el-ph}(T)$ terms refer to phonon-phonon anharmonic and electron-phonon interactions, respectively. In the presence of finite renormalization. When spin-phonon and electron–phonon couplings are absent in a system, the renormalization in phononfrequency occurs predominantly due to the contributions from both the quasi-harmonic ($\Delta\omega^{qh}(T)$) and phonon-phonon anharmonic interactions ($\Delta\omega^{anh}(T)$). The quasi-harmonic term can be quantified using the expression.

$$\frac{\Delta\omega^j}{\omega^j} = \gamma^j \frac{\Delta V}{V} \qquad [3]$$

where $\Delta\omega^j$ represents the quasi-harmonic frequency shift of the phonon mode '$\omega^j$' upon variation of temperature, $\gamma^j$ represents the Grüneisen parameter for the mode 'j' and $\Delta V/V$ signifies the volume change over temperature. In general, the change in unit cell volume due to temperature variation typically amounts to 1-2 % change in the phonon frequency. Considering the typical optical mode Grüneisen parameter $\gamma \sim 1$ and the estimated volume change of ~ 0.7 % (between 90 - 400 K) from our x-ray diffraction data in Eqn. 3, $\frac{\Delta\omega^j}{\omega^j}$ turns out to be ~ 0.7 %. However, the total $\frac{\Delta\omega}{\omega}$ for the phonon mode N1 is observed to be ~ 45 % between 80 and 300 K (see inset in Fig. 2(b) in the main text) and, therefore, the $\frac{\Delta\omega^j}{\omega^j}$ quasi-harmonic contribution is negligible as compared to the total change $\frac{\Delta\omega}{\omega}$ observed over temperature. On the other hand, the sign of $\frac{\Delta\omega}{\omega}$ (for N1, N8, and N9) and $\frac{\Delta\omega^j}{\omega^j}$ (~ 0.7 %) are opposite which implies that the anharmonic term ($\frac{\Delta\omega}{\omega}$) is dominantly high and negative. Therefore, the observed anomalies in phonon behavior of the modes N1, N8, and N9 can be attributed to the intrinsic phonon-phonon anharmonic interactions.

## V. ESTIMATION OF SPIN-PHONON COUPLING CONSTANTS

### A. Experimental Approach

As can be seen in Fig. 2(a) (main text), the phonon frequencies clearly deviate from the anharmonic trend at temperatures below ~ 100 K, which is well above the $T_N$ (~ 8.6 K). As the short-ranged magnetic interactions develop at temperatures well above Ts ~ 35 K, spins simultaneously couple with the phonons leading to the deviation in their behavior from anharmonicity. Conventionally, the Mean-Field approach is employed to estimate the strength ($\lambda$) of spin-phonon coupling (SPC) by evaluating the difference between experimental ($\omega^{exp}(T)$) and the expected anharmonic ($\Delta\omega^{anh}(T)$) frequencies. The observed deviation in frequency from the anharmonic behavior, i.e., $\Delta\omega^{sp-ph}(T) = \omega^{exp}(T) - \omega^{anh}(T)$, can thus be expressed in terms of spin-spin correlation function $<S_i.S_j>$ between the nearest neighbors on two opposite sub-lattices as:

$$\Delta\omega^{sp-ph}(T) = \lambda <S_i.S_j> \propto \sum_r \frac{\partial^2 J}{\partial u_r^2} <S_i.S_j> \quad [4]$$

where, $\frac{\partial^2 J}{\partial u_r^2}$ represents the second-order derivative of magnetic exchange $J$ with respect to the atomic displacement '$u$' along the direction '$r$' [18, 19]. To recall that $Ba_2NiTeO_6$ exhibits a short-ranged ordering well above $T_N$ ~ 8.6 K. Therefore, to take into account its contribution above $T_N$, we make use of the method developed by Cottom and Lockwood which defines $<S_i.S_j>$ in terms of the short-range order parameter $\phi(T)$ for spin S as $<S_i.S_j> = - S^2 \phi(T)$ where $\phi(T) \propto (2T_N/T+T_N)^2$ [20]. It should be noted that as per the Mean-Field approximation, $\phi(T)$ becomes zero above the Néel temperature ($T_N$). However, to include the finite short-ranged spin-correlation above $T_N$, Cottom et al. [20, 21] evaluated $\phi(T)$ for S=1 using the two-spin cluster method. Thus, $\lambda$ can be estimated for S = 1 spin-state of $Ni^{2+}$ at the lowest temperature ($T_{low}$ = 5 K in our case) as [20, 21]:

$$\lambda = -\frac{\Delta\omega^{sp-ph}(T_{low})}{[\Phi(T_{low})-\Phi(2T_N)]\, S^2} \quad [5]$$

where $\Delta\omega^{sp-ph}(T_{low})$ represents the deviation from anharmonic behavior, S refers to spin value, $\phi(T_{low}) = 0.94$ and $\phi(2T_N) = 0.15$ refers to the short-ranged order parameter at temperatures $T_{low} = 5$ K and TN, respectively. The values of φ at required temperatures are obtained from Ref. [20]. A list of estimated coupling coefficients (λ) for different phonons is provided in Table I (main text). To be noted that the phonons N1, N8, and N9 exhibit anomalous behavior with temperature, and cannot be described using the conventional cubic anharmonic equation (Eqn. 1 in the main text). Therefore, at first, we extracted the quasi-harmonic contribution ($\Delta\omega^{qh}(T)$) for these modes using the temperature-dependent volume information obtained from x-ray diffraction (refer to Eqn. 3 in section IV) for temperatures above 100 K (onset of spin-phonon coupling and deviation of $\chi^{-1}$ from Curie law) and extrapolated it down to 5 K. $\Delta\omega^{qh}(T)$ was then subtracted from the experimental ω(T) to obtain:

$$\Delta\omega^{anh}(T) + \Delta\omega^{sp-ph}(T) = \omega(T) - \Delta\omega^{qh}(T) \quad (6)$$

The intrinsic anharmonicity part ($\Delta\omega^{anh}(T)$) of the phonon modes was approximately fitted with the linear temperature dependence (for T > 100 K) and extrapolated down to 5 K which shows a clear deviation below 100 K (see Fig. S3). The difference between the extrapolated curve and $\Delta\omega^{anh}(T) + \Delta\omega^{sp-ph}(T)$ at 5 K, yields the value of λ for N1, N8, and N9 to be 4.8, 1.2, and 1.5 cm$^{-1}$, respectively. The mechanism to mediate spin-phonon coupling in mode N1 is discussed in the main text.

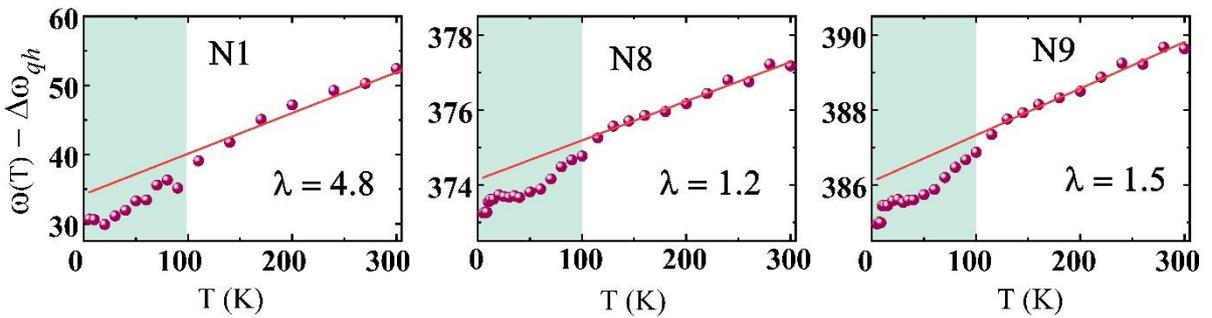

**Fig. S3**: Quasi-harmonic contribution corrected phonon frequencies of the N1, N8, and N9 modes showing deviation from linearity (shaded region), suggesting the spin-phonon coupling below 100 K.

Unlike N1, only oxygen atoms are involved in the vibrations of modes N8 and N9. In the case of N8, it is the stretching motion of oxygen atoms linked to the corner-shared Te-octahedra

(connecting the Ni atoms) that mediates the super-exchange interaction ($J$) through modulation of bond length and angle in $Ni^{2+}$-$O^{2-}$-$Te^{6+}$-$O^{2-}$-$Ni^{2+}$ chain. The atomic chain for N8 is the same as that in N1 but with different eigenvectors. On the other hand, for the N9 mode, the scissoring motion of oxygen atoms linked to face-shared Te-octahedra in $NiO_6$-$TeO_6$-$NiO_6$ trimer unit mediates the spin-phonon coupling through $Ni^{2+}$-$O^{2-}$-$O^{2-}$-$Ni^{2+}$ super-exchange route by modulation of bond length and angles. The spin-phonon coupling strength in $Ba_2NiTeO_6$ is found to be comparable to those in $MnF_2$, $HoCrO_3$, $Ca_2RuO_4$, $Fe_3GeTe_2$, and $Ni_2NbBO_6$ [22–26]. However, a remarkable distinction to note here is that all the examples cited above exhibit a normal super-exchange pathway between two magnetic ions mediated via a single anion as compared to the long super-exchange pathways in $Ba_2NiTeO_6$ with two intermediate ions mediating the magnetic exchange between two $Ni^{2+}$ ions [16, 27]. Therefore, we believe that the spin-phonon coupling in $Ba_2NiTeO_6$ is unique and remarkable.

## B. Theoretical Approach

The estimation of the spin-phonon coupling constants ($\lambda$) from theoretical calculations is made based on the difference in frequencies in the antiferromagnetic (AFM3) and paramagnetic (PM) phases as per the following expression

$$\lambda = \frac{\omega(\text{AFM3}) - \omega(\text{PM})}{<S_i.S_j>} \quad (7)$$

where $\omega(\text{AFM3})$ represents the phonon frequencies of the AFM3 configuration and $\omega(\text{PM})$ represents the frequencies of the paramagnetic phase. Notably, the frequency of the paramagnetic phase in theory, as mentioned in the main text, is considered as an average of the frequencies of the non-magnetic (NM), ferromagnetic (FM), antiferromagnetic-1 (AFM1), antiferromagnetic-2 (AFM2), and antiferromagnetic-3 (AFM3) phases, which are a standard and established approach in the community [28–30]. Moreover, $<S_i.S_j>$ is considered to be 1 as spin S=1 for the $Ni^{2+}$-ion. Since different approaches have been used for the estimation of spin-phonon coupling parameters in experiment and theory, it is likely to give a difference in the values of $\lambda$.

## VI. PHONON ASSIGNMENTS

Raman spectra recorded in the temperature range of 5 - 600 K show 16 phonon modes which are labeled as N1 to N16. The phonon frequencies measured at 5 and 300 K are listed in Table S1. We have also estimated the phonon frequencies using the first-principles calculations for the antiferromagnetic phase (AFM3 configuration as discussed above). The group-theoretically predicted 18 modes ($7A_{1g} + 2A_{2g} + 9E_g$), out of which $A_{2g}$ are silent, have been observed both theoretically and experimentally, majority of which match very well, as shown in Table S1. The symmetries of the corresponding phonons are also assigned in Table S1 below. However, 2 $A_{1g}$ and 1 $E_g$ modes of both experiment and theory do not match well with each other, the reason for which is not clear at present.

**Table S1**: Comparison of experimentally and theoretically obtained frequencies a low- and room-temperature. Additionally, the modes assignment and the atomic species involved in each phonon mode are presented.

| Phonons | Freq. Exp. (300K) (cm$^{-1}$) | Freq. Exp. (5K) (cm$^{-1}$) | Freq. Theo. (0K) (cm$^{-1}$) | Symmetry | Associated atoms |
|---|---|---|---|---|---|
| N1 | 55 | 29 | 31 | $E_g$ | Ba, Ni-tr., O-rock. |
| @ | Silent | Silent | 69 | $A_{2g}$ | O-rotn. |
| N2 | 92 | 93 | 86 | $A_{1g}$ | Ba, Ni-tr., O-scis. |
| N3 | 109 | 109 | 104 | $E_g$ | Ba, Ni-tr., O-rotn. |
| N4 | 121 | 122 | 115 | $A_{1g}$ | Ba-tr., O-rotn. |
| @ | Silent | Silent | 170 | $A_{2g}$ | O-rotn. |
| N5 | 180 | 184 | 171 | $E_g$ | Ba, Ni-tr., TeO$_6$-rotn., O-rotn. |
| N6 | 226 | 232 | 222 | $E_g$ | Ni, O-tr. |
| N7 | 252 | 256 | - | - | - |
| N8 | 378 | 374 | 340 | $E_g$ | O-scis.+rock. |
| N9 | 391 | 386 | 382 | $E_g$ | Ni-tr., O-scis.+wag. |
| N10 | 409 | 410 | - | - | - |
| N11 | 482 | 487 | 440 | $A_{1g}$ | Ni-tr., O-scis.+wag |
| N12 | 581 | 584 | 572 | $E_g$ | Ni-tr., O-sym.+wag.+asym. |
| N13 | 613 | 623 | 597 | $E_g$ | O-sym. + asym. |
| N14 | 684 | 687 | 670 | $A_{1g}$ | O-sym. |
| N15 | 746 | 748 | 704 | $A_{1g}$ | Ni-tr., O-sym. |
| N16 | 752 | 759 | - | - | - |

tr.- Translational, rock.- Rocking, scis. - Scissoring, wag. - Wagging, rotn.- Rotation, sym.- Symmetric, asym.-Asymmetric

Note: @ represents the Silent Raman vibrations

## VII. MAGNETIC CONFIGURATIONS CONSIDERED IN THE CALCULATIONS

In order to elucidate the origin of the unstable phonon mode (N1 at ~55 cm$^{-1}$), different magnetic configurations were considered until a stable frequency was obtained. N1 possesses a negative frequency in the FM state which reduces for both AFM1 and AFM2 phases. Interestingly, N1 becomes positive and stable for the AFM3 state, which is also the expected magnetic ground state for Ba$_2$NiTeO$_6$ as per previous reports [16, 27]. Different magnetic configurations considered in our calculations are shown in Fig. S4.

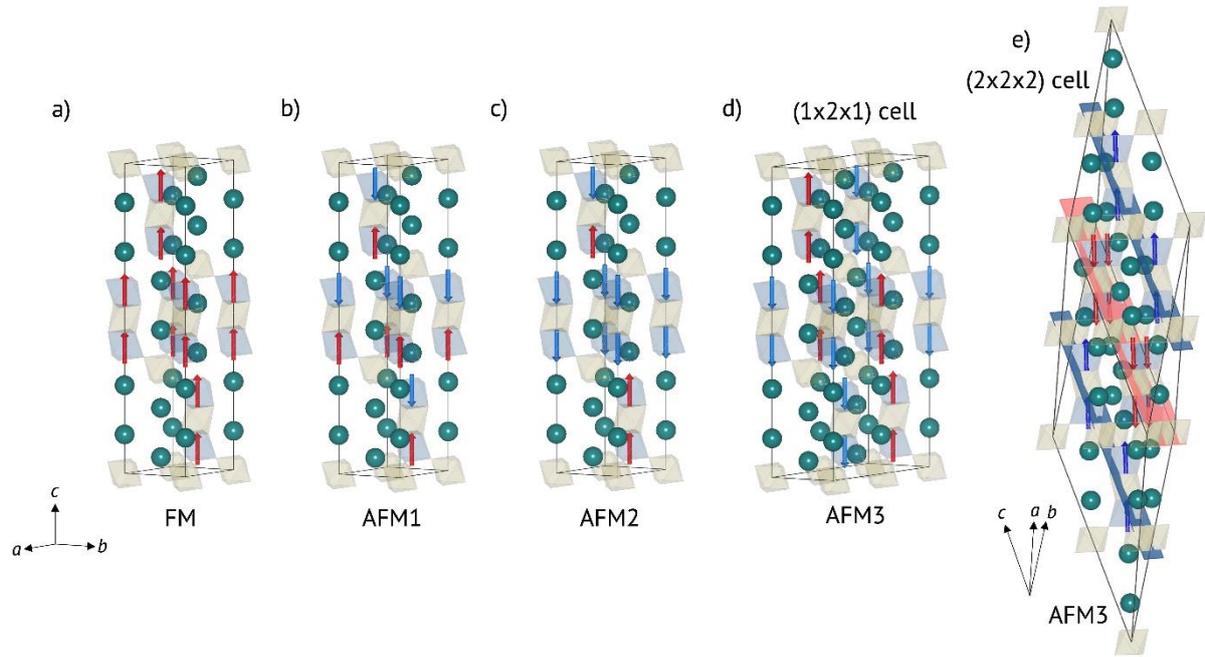

**Fig. S4**: Magnetic orderings of $Ba_2NiTeO_6$ considered and included in the hexagonal unit cell reference. FM, AFM1, AFM2, and AFM3 with respect to the unit cell are presented in a), b), c), and d), respectively. The AFM3 magnetic ordering into the 2×2×2 supercell with respect to the primitive cell is also shown in e). The red and blue arrows represent the up and down magnetic moment orientations. Additionally, the magnetic planes are presented in the AFM3 in red and blue colors, respectively. The Ba-site, $NiO_6$, and $TeO_6$ octahedra are shown, by notation, in dark green, blue, and camel colors, respectively.